\newcommand{\bee}{\begin{equation}}
\newcommand{\ene}{\end{equation}}
\newcommand{\beea}{\begin{eqnarray}}
\newcommand{\enea}{\end{eqnarray}}

\documentclass[aps,prl,showpacs]{revtex4-1}
\usepackage{graphicx}% Include figure files
\usepackage{dcolumn}% Align table columns on decimal point
\usepackage{bm}% bold math
\begin{document}
\title{Phase Transition in Relativistically Degenerate Quantum Plasmas}
\author{M. Akbari-Moghanjoughi}
\affiliation{Azarbajan University of Shahid-Madani, Faculty of Sciences, Department of Physics, 51745-406 Tabriz, Iran}

\begin{abstract}
In this paper we present the generalized phase separation diagram for quantum plasmas revealing the fundamental mechanism called the spontaneous core collapse (SCC) for a white dwarf star core-density based on the extended Shukla-Eliasson force for relativistically degenerate matter taking into account the most important electron interaction features such as Fermi-Dirac statistical pressure, Coulomb, exchange-correlation and quantum tunneling effects. A critical core mass-density value is reported for white dwarf stars beyond which the SCC mechanism will be inevitable. It is revealed that the SCC critical core-density value is slightly affected by the ion charge state, $Z$ (the core composition). Such fundamental effect which is mainly caused by the electron quantum diffraction effect can progressively lead the white dwarf star with a mass beyond the Chandrasekhar limit through SCC, core fusion and the ultimate supernova stages.
\end{abstract}

\maketitle

\section{Introduction}

During the past decade, a new dimension of research has initiated due to Bohmian collective interpretation of quantum systems \cite{1,2,3}. Foundations of quantum hydrodynamic (QHD) formulation, earlier suggested by Madelung \cite{4} termed as hydrodynamic formalism for quantum mechanics, was extended by Takabayasi \cite{5,6,7,8,9} and others. It has been shown that in quantum system of Fermionic particles a new force due to the interaction of particle wavefunction may result in important collective plasma dynamic features. Such ne collective feature is called the Bohm quantum force \cite{10,11} and is the main ingredients of the quantum hydrodynamics (QHD) formulations. In dense plasmas where the inter-electron distances are comparable to or lower than the de Broglie thermal-wavelength $\lambda_D = h/(2\pi m_e k_B T)^{1/2}$, quantum effects such as the Pauli exclusion \cite{12} and the electron quantum Bohm force will be inevitable. Quantum hydrodynamic theory has been successfully employed to interprets the collective dynamics of electron-hole plasmas in a semiconductor \cite{13,14,15}. Variety of applications may be foreseen for quantum plasmas other than semiconductors \cite{16}. For instance, the physical characterization of nanofabricated devices fundamentally rely on the quantum theory of hydrodynamics.

A large number of recent investigations based on QHD model confirm signiﬁcant role of quantum diffraction effect on linear and nonlinear wave dynamics of degenerate plasmas \cite{17,18,19,20,21,22}. It has been shown that the inclusion of Bohm force can lead to many critical features \cite{24,25}. More recently Shukla and Eliasson by investigating the screening effects in a quantum plasma have discovered a new attractive force between ions \cite{26} which is due to interplay between electron degeneracy, exchange-correlation and quantum diffraction effects. The existence of such attractive force among ions, if established experimentally, can mark a distinguished milestone in the history of plasma physics research with the Bohmian collective interpretation of the quantum particles. Moreover, the Wigner-Poisson formulation of hydrodynamics has been recently extended to include the electron spin magnetization effects \cite{27,28} or the relativistic electron momentum effects \cite{29}. Using the quantum magnetohydrodynamic (QMHD) theory, it has been confirmed that the electron spin can play signiﬁcant role in linear and nonlinear wave dynamics of magnetized quantum plasmas \cite{30,31,32,33,34,35}.

One of the important applications of QHD theory regards the astrophysical applications of degenerate plasmas which are ubiquitous in the nature \cite{36,37,38,39,40,41}. Chandrasekhar \cite{42,43} has shown that the compact stars called white dwarfs can be effectively modeled as the ideal zero-temperature quantum plasmas. It has also been shown that the large external gravitational force on stars can lead to a phenomenon known as the relativistic degeneracy \cite{44} leading to distinct features in equation of state (EoS) of degenerate matter \cite{45} and nonlinear wave dynamics \cite{46} in different density regimes. However, the electron diffraction mechanism is supposed to be affected by the relativistic plasma effects, since, the relativistic effects are directly coupled to the plasma number density in the relativistically degenerate matter. Mettes and Sorg \cite{49} and Carlos and Mahecha \cite{50,51,52,53} have extended the fluid dynamics to include the generalized relativistic Bohm force by solving the time dependent Dirac or Klein-Gordon equations. More recently, a relativistic fluid model based on Wigner function and Dirac equation has been proposed \cite{54} which includes the relativistic corrections to quantum diffraction effect.

In the following, we consider a zero-temperature relativistically degenerate collisionless electron/ion fluid with nondegenerate ions of charge state, $Z$. The relativistic quantum hydrodynamic fluid model incorporating the relativistic electron motion, quantum statistical pressure, interaction effects such as Coulomb, exchange-correlation, etc. and the relativistic quantum Bohm force \cite{54}, may be written as

\begin{equation}\label{model}
\begin{array}{l}
\frac{{\partial \gamma n}}{{\partial t}} + \nabla \cdot(\gamma n{\bf{u}}) = 0, \\
\gamma \left( {\frac{{\partial {\bf{u}}}}{{\partial t}} + ({\bf{u}}\cdot\nabla ){\bf{u}}} \right) \\ = \frac{e}{{{m_e}}}\nabla \phi  - \frac{1}{{{m_e}n}}\nabla {P_G} + \frac{{{\hbar ^2}}}{{2\gamma m_e^2}}\nabla \left( {\frac{{\Delta \sqrt n }}{{\sqrt n }}} \right), \\
\Delta \phi ({\bf{r}}) = 4\pi e (\gamma n - {n_0}), \\
\end{array}
\end{equation}

where, $P_G$ is the generalized pressure due to relativistic electron degeneracy in addition to other electron interaction pressures \cite{55}. Also, $\gamma=1/\sqrt{1-\beta^2}$ is the relativistic factor with $\beta=u/c$. It should be noted that the additional correction to the relativistic Bohm force given in Ref. \cite{54} is ignored for being relatively small proportional to $\beta/c$. Using the concept of generalized effective potential, same as in Ref. \cite{56}, one may obtain the plasma dielectric response to harmonic perturbations via the Fourier analysis of Eqs. (\ref{model}) as follows

\begin{equation}\label{eff}
\begin{array}{l}
- \omega {n_1} + {\bf{k}}\cdot{{\bf{u}}_{\bf{1}}} = {\rm{0}}, \\
- \omega {{\bf{u}}_{\bf{1}}} =  - \frac{{{\hbar ^2}}}{{4m_e^2}}{k^2}{\bf{k}}{n_1} - L{c^2}{\bf{k}}{n_1},\\
L = \frac{{{R_0^2}}}{{3\sqrt {1 + {R_0^2}} }} - \frac{\beta}{3} R_0 + \frac{a_0 }{{2\pi }}\left[ {\frac{R_0}{3} - \frac{R_0}{{1 + {R_0^2}}} + \frac{{{R_0^2}{{\sinh }^{ - 1}}R_0}}{{{{(1 + {R_0^2})}^{3/2}}}}} \right],
\end{array}
\end{equation}

It is evident that the relativistic parameter, $\gamma$, does not enter the linear response model. In the linearized model, Eq. (\ref{eff}), the parameter $R_0=(n_0/n_c)^{1/3}$ with $n_c=5.9\times 10^{29}cm^{-3}$ is the Chandrasekhar's relativity parameter, $a_0=e^2/\hbar c\simeq 1/137$ is the fine-structure constant and $\beta=(2{a_0}/5){(2Z/3\sqrt \pi  )^{2/3}}$. Therefore the static dielectric constant is then given as

\begin{equation}\label{di}
D(0,{\bf{k}}) = 1 + \omega _{pe}^2{\left[ {\frac{{{\hbar ^2}{k^4}}}{{4m_e^2}} + L{c^2}} \right]^{ - 1}},
\end{equation}

where $\omega_{pe}=\sqrt{4\pi n_0 e^2/m_e}$ is the electron plasmon frequency. The dielectric around a point-like stationary charge, $Q=Ze$, located at the origin is given by

\begin{equation}\label{phi}
\phi ({\bf{r}}) = {\frac{Q}{{2\pi^2 }}}\int {\frac{{\exp (i{\bf{k}} \cdot {\bf{r}})}}{{{k^2}D(0,{\bf{k}})}}{d^3}{{k}}},
\end{equation}

Decomposition of the integral with known solution given in \cite{26} reads as

\begin{equation}\label{sol}
\phi (r) = \frac{Q}{{2 r}}\left[ {\left( {1 + b } \right){e^{ - {k_-}r}} + \left( {1 - b } \right){e^{ - {k_+}r}}} \right].
\end{equation}

with the potential parameters defined as below

\begin{equation}\label{par}
\begin{array}{l}
k_{\mp}^2 = k_0^2\frac{{1 \mp \sqrt {1 - 4\alpha } }}{{2\alpha }},\hspace{3mm} b = \sqrt {1 - 4\alpha },\\ \alpha  = \frac{{{\hbar ^2}\omega _{pe}^2}}{{4{m_e^2}c^4{L^2}}},\hspace{3mm}{k_0} = \frac{{{\omega _{pe}}}}{{c\sqrt L }}.
\end{array}
\end{equation}

One of the fundamental characteristics of the Shukla-Eliasson potential \cite{26} given by Eq. (\ref{sol}) is that for $\alpha>1/4$ the potential is attractive and repulsive otherwise. The generalized degenerate matter phase separation then is obtained by evaluation of the parameter, $\alpha$. In Fig. 1 we have shown the variation of this parameter for wide density parameter, $r_s=r_B/r_0$ range, where, $r_B=\hbar^2/m_e e^2\simeq 5.3\times 10^{-8}cm$ is the Bohr-radius and $r_0=(3n_0/4\pi)^{1/3}$ is the Wigner-Seitz cell radius defined as a sphere containing only a single electron. The relativity parameter is given as $R_0\simeq 0.014r_s$.

\begin{figure}[htb]
\centering
\includegraphics[width=8cm]{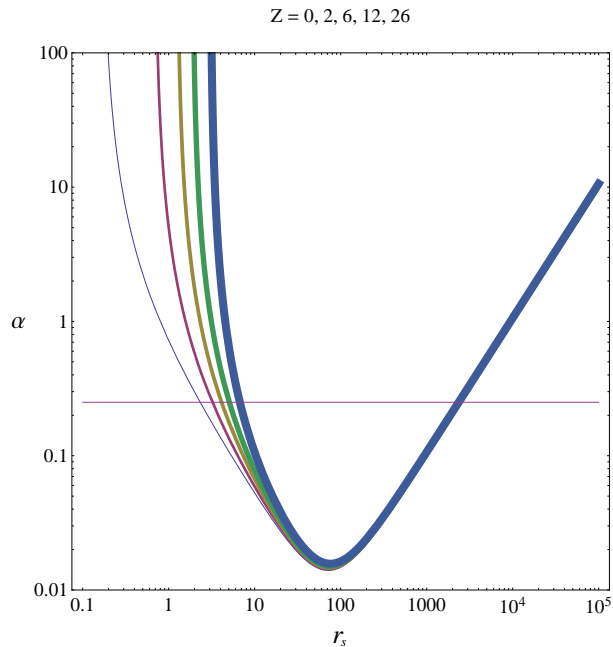}
\caption{The variation of $\alpha$ against the plasma density parameter, $r_s=a_B/r_0$. The critical value $\alpha=1/4$ is indicated
with the horizontal line. The thickness of curves appropriately reflect the variation in the ion charge value, $Z$}
\end{figure}

\begin{figure}[htb]
\centering
\includegraphics[width=15cm]{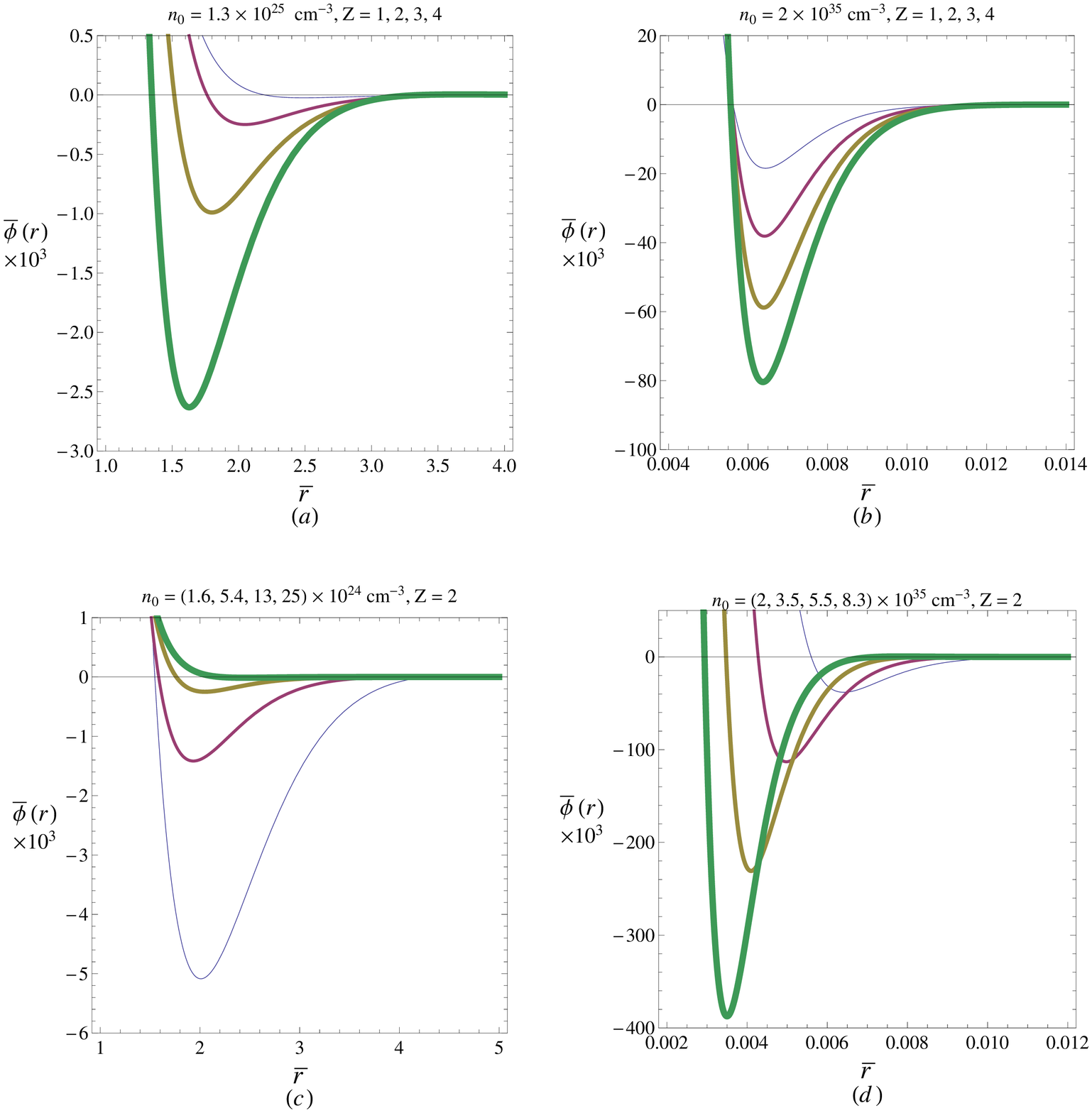}
\caption{(Color online) Variations of normalized electric potential $\bar\phi=r_B\phi/Ze$ for $\alpha>1/4$ with changes in plasma density and the ion charge-state for nonrelativistic (Figs. a and c) and relativistic (Figs. b and d) degeneracy density regimes. The distance $\bar r=r/r_B$ is measured in the Bohr-radius unit. The thickness of curves appropriately reflect the variation in the varied plasma parameter, in each plot.}
\end{figure}

As it is observed form Fig. 1, increasing $r_s$ (increasing the plasma density) from zero-density first leads the pure Coulomb screening ($\alpha=+\infty$) to turn over to the Lennard-Jonse-like Shukla-Eliasson attractive potential of form $\phi (r)=(Ze/r) \exp[ - \Re ({k_-})r]\{ \cos [\Im({k_-})r] + b\sin [\Im ({k_-})r]\}$, where $\Re$ and $\Im$ refer to the real and imaginary parts of the parameter $k_-$. For two different values of $r_{s1}(Z)$ and $r_{s2}(Z)$ we have $\alpha=1/4$. For $r_{s1}(Z)<r_s<r_{s2}(Z)$ the plasma undergoes the ordinary Debye-H\"{u}ckel screening and for $r_s>r_{s2}(Z)$ the Shukla-Eliasson attractive force reappears. It is remarked that, the value of $r_{s2}(Z)$ unlike that of $r_{s1}(Z)$ is not strongly sensitive to the ion charge state, $Z$.This value coincides exactly with the core-density of a typical white dwarf star ($10^{8}<\rho_{c}(gr/cm^3)<10^{13}$). It is remarked that $r_{s1}=6.8326$ value for iron plasma composition extends to density of $\rho\simeq 821.945 gr/cm^3$ which is well beyond the zero-pressure metallic iron. This is the indication that the Shukla-Eliasson attractive potential leading to the plasma crystallization can be present at small-planet core density ranges. On the other hand, the $r_{s2}\simeq 2354.8$ value for the same plasma composition corresponds to the mass density of $\rho\simeq 3.5\times 10^{10}gr/cm^3$ which resides in the range of a typical white dwarf core. The reappearance of Shukla-Eliasson attractive force in white-dwarf core density has some fundamental consequences for stellar evolution, as will be discussed next.

Figure 2 reveals some fundamental differences of Shukla-Eliassom attraction between nonrelativistic and relativistic degeneracy cases. Figures 2(a) and 2(b) compare the potential profiles (normalized binding energy, $r_B\bar\phi(r)\phi/Ze$) in nonrelativisic and relativistic degeneracy cases against the variation in the ion number density, $Z$, where $\bar r=r/r_B$. It is revealed that, the increase in the number-density leads to increase in the potential depth in both degeneracy regimes while this variation is more pronounced in the nonrelativistic degeneracy case. However, as it is remarked comparing Figs. 2(c) and 2(d) the increase in the plasma density causes reverse effects in the two density regimes. In other words, the attractive potential strength (binding energy) increases/decreases for the case of relativistic/nonrelativistic degeneracy regimes due to increase in the plasma density. The variation in the potential-minimum location also follows quite dissimilar rout in the two degeneracy regimes.

\begin{figure}[htb]
\centering
\includegraphics[width=15cm]{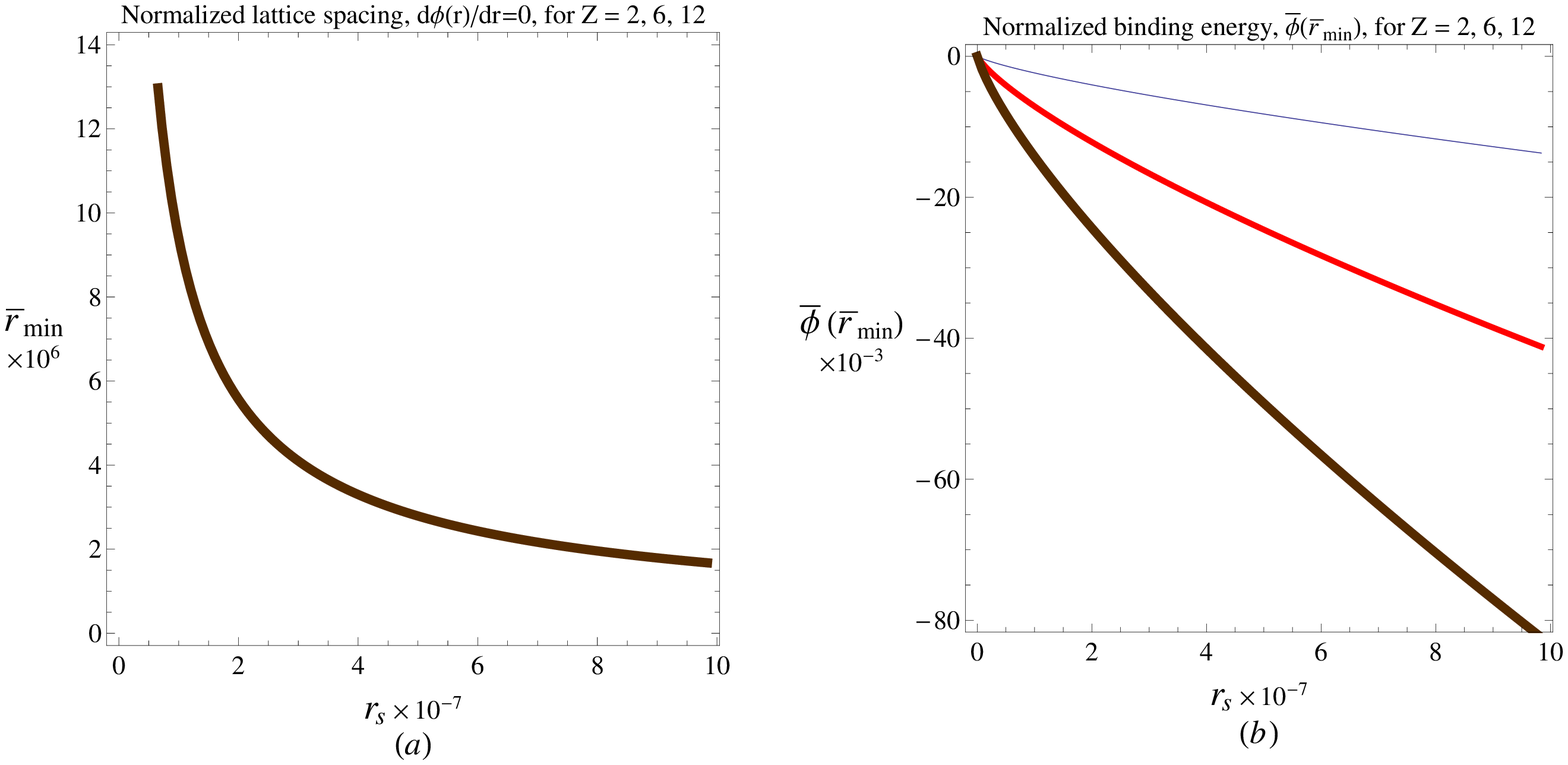}
\caption{(Color online) Variation in the binding energy and length in the relativistic degeneracy regime for three different plasma compositions, namely, Helium, Carbon and Magnesium. The thickness of curves being a measure in the value of the varied parameter in each plot.}
\end{figure}

Figure 3 closely inspects the variations of Shukla-Eliasson attractive potential parameters, namely, the binding energy and binding length for the relativistic degeneracy regime relevant to the white dwarf core density for three different core compositions of Helium, Carbon and Magnesium. It is observed that the unlike for the binding energy, $\bar\phi(\bar r_{min})$, the binding length, $\bar r_{min}$, is unaffected by the change in plasma atomic-number, $Z$. It is further remarked that in the relativistic densities the binding length decreases suddenly with increase of the plasma density and reaches an asymptotic value of about $10^{-14} cm$, which is which is two orders of magnitude smaller that the nucleon diameter which may seem controversial. However, One should note that the hydrodynamic theory used in this model may not even be valid beyond the quantum screening length. On the other hand, other processes like nuclear fusion takes place before such short ionic distances are met. Figure 3(b) within a fixed relativistic density heavier atoms crystallize into  stronger lattices. Crystallization in white dwarf cores are sometimes wrongly attributed to the presence of strong gravitational inward pressure at the center of the star. However, such confinement of ions in the core may only lead to quasi-lattice formation with viscous ion fluid due to lowering of ion coupling parameter value in fusion-medium temperatures. Slattery {\it et al.} \cite{57} have shown that for ion coupling parameter values of $\Gamma<178$ the plasma will be in liquid state. Therefore, Shukla-Eliasson attractive force can attribute appropriately to the perfect crystallization in a wide density range relevant to both cores of small planets as well as that of a typical white dwarf star.

\begin{figure}[htb]
\centering
\includegraphics[width=8.5cm]{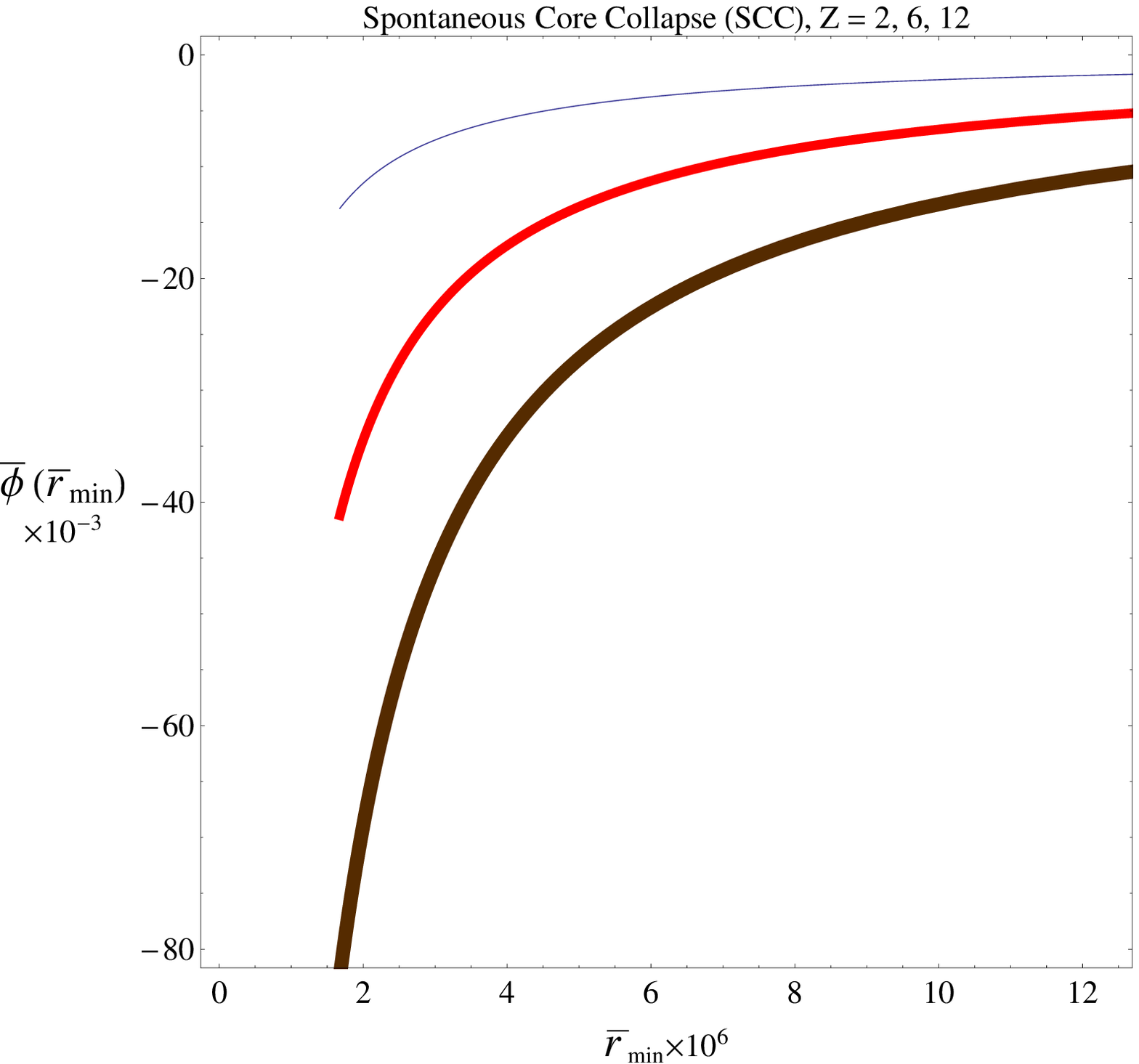}
\caption{(Color online) Variation of the normalized plasma binding energy in terms of the binding length for various plasma composition, namely, Helium, Carbon and Magnesium. The plot shows a feature specific to relativistic degeneracy regime called the spontaneous core collapse relevant to white dwarf stars. The thickness of curves being a measure of the value of the varied parameter.}
\end{figure}

Figure 4 reveals yet another fundamental feature of Shukla-Eliasson screening for relativistic density regimes. It depicts variation of the normalized binding strength versus the normalized binding length for different plasma compositions. It is observed that with decrease of the binding length the binding energy increases for all plasma composition with more pronounced increase for higher plasma number density. Such variations reveal that the plasma crystals with relativistic density are unstable under external compression. Any force exerted from outside such as the gravitational inward pull for the case of a compact star would decrease the binding length and lead to collapse of the configuration. Such progressive increase in the plasma binding energy due to the gravitational pressure would also result in the energy release in the core leading the nuclear fusion ignition. As the atoms fuse they will form heavier elements which are lead to form even more stronger bounds and the release of more energy. Such loop-hole may proceed infinitely until the ultimate collapse of the white dwarf core releasing enough energy to lead to a cataclysmic supernova to take place. It should be noted that such mechanism called the spontaneous core collapse is only possible in the relativistic degeneracy regime, $r_s>r_{s2}(Z)$ and is not possible for $r_s<r_{s1}(Z)$.

We conclude that Shukla-Eliasson quantum screening potential which goes beyond the ordinary Debye or the Thomas-Fermi models can successfully explain the structure formation in solid-density matter, cores of small planets and compact stellar objects. Such theory which relies on the complete quantum interpretation of collective fermion behavior can even lead to deep understanding of the stellar formation and structure. It can also result in breakthroughs in pure as well rapidly growing nanotechnology sciences.

\end{document}